\documentclass[superscriptaddress,altaffilletter,amsmath,amssymb,aps,twocolumn,floatfix]{revtex4-1}
\usepackage[english]{babel}
\usepackage{multirow}
\usepackage{comment}
\usepackage{float}
\usepackage{textcomp}
\usepackage[justification=justified]{subcaption}
\usepackage[justification=justified]{caption}
\usepackage[compact]{titlesec}
\titlespacing{\section}{0pt}{3ex}{2ex}
\titlespacing{\subsection}{0pt}{2ex}{2ex}
\titlespacing{\subsubsection}{0pt}{1.5ex}{1.5ex}

\usepackage{graphicx}
\usepackage{epstopdf}
\usepackage{color}
\usepackage{dcolumn}
\usepackage{bm}
\usepackage{hyperref}
\usepackage[mathlines]{lineno}
\usepackage{upgreek} 
\usepackage{threeparttable} 
\usepackage{booktabs} 
\usepackage{siunitx}

\begin{document}

\title{Identification of Radiopure Titanium for the LZ Dark Matter Experiment and Future Rare Event Searches}  
\author{D.S.~Akerib}
\affiliation{SLAC National Accelerator Laboratory, P.O. Box 20450, Stanford, CA 94309-0450, USA}
\affiliation{Kavli Institute for Particle Astrophysics and Cosmology (KIPAC), Stanford University, 452 Lomita Mall, Stanford, CA 94309-4008, USA}

\author{C.W.~Akerlof}
\affiliation{University of Michigan, Randall Laboratory of Physics, 450 Church Street, Ann Arbor, MI 48109-1040, USA}

\author{D.~Yu.~Akimov}
\affiliation{National Research Nuclear University MEPhI (NRNU MEPhI), 31 Kashirskoe shosse, Moscow, 115409, Russia}

\author{S.K.~Alsum}
\affiliation{University of Wisconsin-Madison, Department of Physics, 1150 University Avenue Room 2320, Chamberlin Hall, Madison, WI 53706-1390, USA}

\author{H.M.~Ara\'{u}jo}
\affiliation{Imperial College London, Physics Department, Blackett Laboratory, Prince Consort Road, London, SW7 2BW, UK}

\author{I.J.~Arnquist}
\altaffiliation{Pacific Northwest National Laboratory, P.O. Box 999, Richland, WA 99352-0999, USA; non-LZ Collaboration}
\noaffiliation

\author{M.~Arthurs}
\affiliation{University of Michigan, Randall Laboratory of Physics, 450 Church Street, Ann Arbor, MI 48109-1040, USA}

\author{X.~Bai}
\affiliation{South Dakota School of Mines and Technology, 501 East Saint Joseph Street, Rapid City, SD 57701-3901, USA}

\author{A.J.~Bailey}
\altaffiliation[Now at: ]{Instituto de Física Corpuscular (IFIC), University of Valencia and CSIC, Parque Científico, C/Catedr\'{a}tico Jos\'{e}, Beltr\'{a}n, 2, E-46980 Paterna, Spain}
\affiliation{Imperial College London, Physics Department, Blackett Laboratory, Prince Consort Road, London, SW7 2BW, UK}

\author{J.~Balajthy}
\affiliation{ University of Maryland, Department of Physics, College Park, MD 20742-4111, USA}

% ...

\collaboration{The LUX-ZEPLIN (LZ) Collaboration}

    \date{\today}
\begin{abstract}
\noindent 
The LUX-ZEPLIN (LZ) experiment will search for dark matter particle interactions with a  detector containing a total of 10 tonnes of liquid xenon within a double-vessel cryostat. The large mass and proximity of the cryostat to the active detector volume demand the use of material with extremely low intrinsic radioactivity.  We report on the radioassay campaign conducted to identify suitable metals, the determination of factors limiting radiopure production, and the selection of titanium for construction of the LZ cryostat and other detector components. This titanium has been measured with activities of $^{238}$U$_{e}$~$<$1.6~mBq/kg, $^{238}$U$_{l}$~$<$0.09~mBq/kg, $^{232}$Th$_{e}$~$=0.28\pm 0.03$~mBq/kg, $^{232}$Th$_{l}$~$=0.25\pm 0.02$~mBq/kg, $^{40}$K~$<$0.54~mBq/kg, and $^{60}$Co~$<$0.02~mBq/kg (68\% CL).  Such low intrinsic activities, which are some of the lowest ever reported for titanium, enable its use for future dark matter and other rare event searches. Monte Carlo simulations have been performed to assess the expected background contribution from the LZ cryostat with this radioactivity. In 1,000 days of WIMP search exposure of a 5.6-tonne fiducial mass,  the cryostat will contribute only a mean background of $0.160\pm0.001$(stat)$\pm0.030$(sys) counts.
\end{abstract}

\maketitle

\section{\label{sec:Introduction}Introduction}
\noindent Direct dark matter searches attempt to measure the interaction of Weakly Interacting Massive Particles (WIMPs) with an atomic nucleus, a process not yet observed. The very nature of this low-energy rare-event search demands extremely low background event rates in the signal region of interest. In recent leading experimental searches, the background has been dominated by the decay of naturally occurring radioactive contamination present within the materials used to construct the detectors, primarily  $^{238}$U, $^{232}$Th, $^{40}$K, and $^{60}$Co. To achieve the design sensitivity, the control of such radioactivity is fundamental not only to direct dark matter experiments, but to rare event searches in general, requiring extensive radioassay campaigns to select suitable construction materials.

LZ (LUX-ZEPLIN) is a second generation dark matter experiment, presently under construction~\cite{akerib:2015cja}. It will feature a two-phase xenon time projection chamber (TPC), containing approximately 7 tonnes of active liquid xenon (LXe)~\cite{Chepel:2012sj}. The total load of 10 tonnes of liquid xenon and other detector components will be held in a cryostat made from two nested vessels---the inner vessel contains the cold xenon whilst the outer vessel enables a vacuum region for thermal insulation. The material from which the cryostat is fabricated must be suitable for operation at cryogenic temperatures, must have mechanical properties that comply with international codes regulating the materials for pressure vessel fabrication, must not be prohibitively expensive and must be extremely radiopure. 

Historically, dark matter searches have used cryostats made from stainless steel or electroformed oxygen-free high conductivity (OFHC) copper.  Stainless steel is inexpensive, readily satisfies mechanical constraints with its high tensile strength, and vessels are straightforward to manufacture. However, $^{238}$U and $^{232}$Th content can be significant and highly variable and levels of $^{40}$K and $^{60}$Co are typically high. Copper is usually found to be considerably more radiopure, but thicker vessels are required relative to stainless steel when manufacturing pressure vessels for equivalent mechanical strength. The  ZEPLIN-III experiment, one of the predecessors to LZ, used a low-activity copper cryostat  containing 12 kilograms of LXe~\cite{lebedenko:2009xe,Akimov:2011tj}. For larger experiments at the tonne-scale and beyond, thick cryostats not only impact total radioactivity, but also attenuate internally-produced $\gamma$-ray and neutron backgrounds, acting to limit the effectiveness of veto detectors that may surround the primary target volume. 

Titanium has previously been considered as an attractive alternative to stainless steel and copper.  It has a high strength-to-weight ratio, is resistant to corrosion, and is already popular in the aerospace, metal finishing, oil refining and medical industries. It has a lower atomic number and therefore lower attenuation of low energy $\gamma$-radiation than stainless steel, should in principle contain very little $^{60}$Co, and also benefits from little cosmogenic activation. This is significant since radioactive isotopes can be produced by neutron-induced spallation reactions whilst the material is still above ground. Stainless steel contains isotopes of iron, nickel and chromium that can be activated to form radioisotopes such as $^{56}$Co, $^{58}$Co, $^{54}$Mn and $^{48}$V~\cite{maneschg:2008zz}, whilst in titanium, only the production of $^{46}$Sc is a concern~\cite{Zhang:2016rlz}.  Although the LUX experiment~\cite{akerib:2013tjd,akerib:2015rjg,Akerib:2016lao,akerib:2016vxi}, LZ's most recent predecessor, successfully manufactured and deployed a cryostat constructed from low-activity titanium to contain 350~kg of LXe~\cite{akerib:2011rr}, securing similarly radiopure material has proved non-trivial across the field of rare-event search experiments.
The importance of constructing vessels of sufficient radiopurity has already been recognized as necessary for the next generation of neutrinoless double-beta-decay (0$\nu\beta\beta$) experiments~\cite{ndbd:2015}.

To procure titanium for the manufacture of the LZ cryostat of similar or better radiopurity to that in LUX, an extensive R\&D campaign has been conducted. This was necessary to ensure that reliably radiopure material could be sourced in sufficient quantities for LZ. The campaign has determined the requirements for procurement of reproducible radiopure titanium metal, included an extensive assay campaign of titanium samples (alongside stainless steel samples as an alternative) and, finally,  selected suitable candidate materials. The measured radioactivities of screened samples have been used as input for detailed studies employing Monte Carlo simulations to model the LZ cryostat within the entire experiment. The impact of the intrinsic contamination in terms of background contribution to the WIMP search is determined, and this informs the selection of the most suitable sample for further confirmatory screening before procurement of the batch material for LZ. 

\section{Titanium Production and Contaminants}

Titanium (Ti), a lightweight transition metal with atomic number 22, average atomic weight 47.90 and a density of 4510 kg/m$^3$, is the fourth most abundant metal on Earth, comprising 0.62\% of the Earth's crust. Titanium production begins with the mining and refinement of ores, and involves exposure to coke, coal, oil and tar which can introduce significant contamination of radioactive elements including uranium and thorium. However, titanium oxide ores are treated with chlorine gas to produce titanium tetrachloride (TiCl$_4$), which is found to have extremely low radiocontent after chemical treatment and filtering~\cite{mozhevitina:2015jla}. 

To produce metallic titanium, the TiCl$_4$ is reduced by liquid magnesium or sodium in a vessel filled only with argon to avoid oxygen or nitrogen contamination. This produces a porous titanium sponge, which is purified by leaching or heated vacuum distillation. There are typically no controls in place at this stage to limit contamination from surfaces or fluids in contact with the sponge, and thus variability in radioactivity of sponge products is expected.

Before melting the sponge, any alloys and scrap metals required are added by mechanical compaction. The use of these, where radioisotope contamination of source material is not traceable, constitutes the highest risk for introduction of radioactivity into the bulk material. At this stage the radiopurity may be greatly affected even if the purity has been preserved from the TiCl$_4$ through to the sponge.

Refinement of titanium sponge can be undertaken using either vacuum arc remelting (VAR) or  electron beam cold hearth (EBCH) techniques. VAR is used for most commercial titanium and involves melting the metal under a vacuum in a crucible. A current is passed across a gap between an electrode consisting of compacted sponge and alloy and a smaller piece of the metal. This creates an arc and begins a continuous melting process. As the electrode melts, it must be lowered towards the bottom of the vessel, allowing control over the solidification rate, which affects the micro-structure of the titanium. The vacuum conditions allow any remaining gas contamination (e.g. N$_2$, O$_2$, H$_2$) to escape the titanium into the vacuum chamber.  The resulting ingot is usually melted either once or twice more. 
Other impurities such as carbon, sulphur and magnesium that have a high vapour pressure will also be reduced in concentration. The disadvantages of VAR are that it is a slow process and nitrogen-rich low density particles survive as their residence time at sufficient temperatures is not long enough to melt them. 

Electron beam remelting uses electron beams to heat and melt metals in a water-cooled copper crucible under a vacuum. In EBCH, cold hearth refers to a watercooled copper hearth in which the feedstock is drip melted, overflowing into a withdrawal mold.  EBCH holds some advantages for manufacturers: it is a faster process as it involves fewer steps than traditional refinement implying fewer opportunities for further radiological contamination, its high yield, its flexibility in melting feedstock of any geometry and finally the possibility of producing ingots and slabs with a wide range of cross sections. A final and key advantage for applications such as low-background experiments is that EBCH exceeds other refinement methods in removing high density contaminants, as these sink to the bottom of the hearth and so do not reach the mold \cite{ASM:2008}.  

As part of the radioassay campaign for LZ, titanium of varying grades has been sampled in partnership with several manufacturers and at various stages of the production cycle beyond radiopure TiCl$_{4}$ (including sponge, slabs, and sheet), using both re-melting processes (VAR and EBCH), and with varying scrap concentrations (with the majority requested and delivered with 0\% scrap). 

\section{\label{sec:Radiassay}Radioassay Campaign}
\subsection{\label{subsec:Screening}Screening Techniques}

The cryostat material assays have been primarily carried out by the Berkeley Low Background Facility using its surface screening detectors at the Lawrence Berkeley National Laboratory and the underground HPGe system \textsc{Maeve}. Until November 2015, \textsc{Maeve} was located underground ($\sim$550~ft) at the Oroville Dam in Oroville, California, but has now been relocated to the Black Hills State University Underground Campus (BHUC) facility dedicated to low-background counting 4,850 ft underground at the Sanford Underground Research Facility (SURF) in Lead, South Dakota. \textsc{Maeve} is an ORTEC HPGe 85\% p-type detector in a low background cryostat, shielded by OFHC copper and lead. For samples of a few kg, the sensitivity after two weeks of counting reaches approximately 10~ppt  U 
($\approx$0.1~mBq/kg $^{238}$U) and 25~ppt Th ($\approx$0.1~mBq/kg $^{232}$Th), 20~ppb for K ($\approx$0.7~mBq/kg $^{40}$K) and $\approx$0.03~mBq/kg  for $^{60}$Co. 
\textsc{Maeve} counting efficiencies were modelled with a \textsc{Geant4} \cite{Agostinelli:2002hh} simulation including precise geometry, accounting for sample dimensions and placement within the detector castle. The GEANT4 models were carefully verified with well-characterized, homogeneous, weakly radioactive sources to ensure the validity of the Monte Carlo simulations for both geometric placement, self-attenuation, and energy response of the HPGe crystal. 

Measurements have also been performed at the Boulby Underground Laboratory in the UK, primarily using \textsc{Chaloner}, a 0.8~kg Canberra BE5030 Broad Energy Germanium (BEGe) detector. This instrument, with the  advantage of its low energy threshold ($\approx$15~keV), was used to perform confirmation measurements, particularly of the low energy and low branching ratio lines close to the top of the $^{238}$U and $^{232}$Th decay chains. Low background HPGe detectors operated at the University of Alabama in a surface laboratory have also been utilized for this campaign. All detectors have been cross-calibrated with results from the assay of standard sources, and agree within stated systematic uncertainties. 

All signal peaks detected with HPGe are passed for analysis, and multiple lines are measured to infer $^{238}$U and $^{232}$Th concentrations, as well as the lines from $^{60}$Co, $^{137}$Cs and other long-lived species. For $^{238}$U it is standard to  measure the decays of $^{226}$Ra, $^{214}$Pb and $^{214}$Bi, and for $^{232}$Th the decays of $^{228}$Ac, $^{212}$Bi, $^{212}$Pb and $^{208}$Tl. Rather than combining these lines to report single values for $^{238}$U and $^{232}$Th, we separately extract both early chain and late chain activities as manufacturing/chemical processes may break secular equilibrium in the chains. This separation is necessary given the different energies and branching ratios of $\gamma$-rays throughout the chains, and more significantly, the impact on neutron production yields through both ($\alpha$,$n$) processes and spontaneous fission. In the case of $^{238}$U, we define the early part of the chain ($^{238}$U$_{e}$) as containing any isotopes above $^{226}$Ra ($T_{1/2}$ = 1600 yr).  The late part of the chain ($^{238}$U$_{l}$) is counted from $^{226}$Ra and below. Another break in the U chain often occurs at $^{210}$Pb ($T_{1/2}$ = 22 yr) and is also reported separately when available.
Among the $^{238}$U sub-chains, contributions from $^{238}$U$_{l}$ dominate LZ $\gamma$-ray and ($\alpha$,$n$) backgrounds~\cite{malling:2013jya}, whilst $\gamma$-rays and neutrons from fission of $^{238}$U in $^{238}$U$_{e}$ are efficiently vetoed~\cite{shaw:2016}.
For the $^{232}$Th chain, we define the early part of the chain ($^{232}$Th$_{e}$) as coming from isotopes above $^{228}$Th and the late part of the chain ($^{232}$Th$_{l}$) as coming from isotopes from $^{228}$Th and below.
Among the $^{232}$Th sub-chains, roughly equal contributions to LZ $\gamma$-ray background arise from $^{232}$Th$_{e}$ and $^{232}$Th$_{l}$, whilst ($\alpha$,$n$) backgrounds arise dominantly from $^{232}$Th$_{l}$.
  
Small samples of titanium have been subjected to direct mass spectrometry for their U and Th content using Inductively Coupled Plasma Mass Spectrometry (ICP-MS) at a dedicated facility set up at University College London (UCL) for LZ assays~\cite{Dobson:2017}, and at the Pacific Northwest National Laboratory (PNNL). Both facilities operated an Agilent 7900 ICP-MS instrument in cleanrooms, taking extreme care to avoid trace contamination of the samples in handling, preparation and assay.

\subsection{\label{subsec:RadioassayResults}Radioassay Results}
Titanium samples were secured for radioassay from several international  manufacturers and suppliers, including VSMPO~\cite{VSMPO:2015}, TIMET~\cite{TIMET:2015}, Supra Alloy~\cite{SupraAlloys:2015}, Honeywell~\cite{HoneywellEM:2015} and PTG~\cite{PTG:2015}.  The samples were collected from stages throughout the manufacturing process with a focus on final processed sheet. TIMETAL\textsuperscript{\textregistered}35A sheet samples, equivalent to ASTM Grade 1, showed the highest radiopurity in early measurements and so several were obtained. After rejection of samples with known high scrap content or where traceability of the sample through the manufacture process was compromised, 22 samples were selected for assay with the high sensitivity underground gamma spectroscopy detectors. The sample masses varied from approximately 0.5~kg to 10~kg and were cleaned prior to assay by etching with hydrofluoric and nitric acids. Several different types of titanium were measured: ASTM Grade 1  which is the highest grade purity of unalloyed titanium with a maximum oxygen content of 0.18\%, ASTM Grade 2  which is also unalloyed but may contain a maximum oxygen content of up to 0.25\%,  porous titanium sponge created during the first stage of processing, and some finished nuts and bolts. The radioassay results from each of these samples are presented in Table~\ref{tab:ti-samples}.

\begin{table*}[t]
\footnotesize
\centering
\caption{\label{tab:ti-samples} \small  Radioassay results of 22 titanium samples (68\% CL) from a variety of suppliers. Numbers in italics are $1\sigma$ upper limits, and $^{210}$Pb results are given where available. Systematic uncertainties are estimated to be up to 10\% by comparing the cross-calibration results of the assay of standard sources. TIMET sponges were obtained as a dirty gravel; the high $^{210}$Pb activities observed are expected due to surface contamination and identify them as unsuitable.}
\begin{tabular}{|l|l|c|c|c|c|c|c|} \hline
 \multirow{2}{*}{\bf{Name}} & \multirow{2}{*}{\bf{Type}} & \multicolumn{3}{|c|}{\bf{$^{238}$U (mBq/kg)}} & \multicolumn{2}{|c|}{\bf{$^{232}$Th (mBq/kg)}}& \bf{$^{40}$K} \\
& &   \bf{early} &  \bf{late} &  \bf$^{210}$Pb &  \bf{early} &  \bf{late} &  \bf(mBq/kg) \\ \hline 
Supra Alloy Sheet (1) & ASTM Grade 1 Sheet  & 32 & 4.2 & - & 3.3 & 2.8 & \it{$<$1.9} \\ \hline
Supra Alloy Sheet (2) & ASTM Grade 2 Sheet  & 110  & \it$<$2 &- &  200 & 180 & 25 \\  \hline
TIMET Sponge (1)  & Sponge  & 25 & \it{$<$2} & 250 & \it$<$4.1 & \it$<$4.1 & \it$<$12 \\ \hline
TIMET Sponge (2) & Sponge  & \it$<$25 & \it{$<$2} & 6200 & \it$<$4.1 & \it$<$2.4 & \it$<$15 \\ \hline
TIMET Sponge (3) & Sponge & \it$<$25 & \it{$<$2} & \it$<$62 & \it$<$5.3 & \it$<$1.6 & \it$<$12 \\ \hline
TIMET Sponge (4) & Sponge  & 74& \it{$<$2} & 120 & \it$<$4.1 & \it$<$1.6 & \it$<$12 \\ \hline
TIMET Sponge (5) & Sponge  & \it$<$12& \it{$<$2} &  740 & \it$<$4.1 & \it$<$1.6 & \it$<$12 \\ \hline
TIMET Sponge (6)  & Sponge  & 74  & \it$<$4 &  2500 &  \it$<$5.3 & 14 & \it$<$19 \\ \hline
TIMET Sponge (7)  & Sponge  & 37 & 25 & 2500 & 12 & 5.7 & \it$<$12 \\ \hline
TIMET Sheet (1) & ASTM Grade 1 Sheet  & 11 & \it$<$0.62 &- & \it$<$0.8 & \it$<$0.6 & \it$<$2.5 \\ \hline
TIMET Sheet (2) & ASTM Grade 1 Sheet  & 5 & 3.3 & - & 2.8 & 0.8 & \it$<$1.5 \\ \hline 
TIMET Sheet (3) & ASTM Grade 1 Sheet  & 8.5 & 0.37 & - & 0.45 & 0.61 & \it$<$0.5 \\ \hline
TIMET Sheet (4) & ASTM Grade 1 Sheet  & 8.0 &  \it$<$0.12 & - & \it$<$0.12& \it$<$0.1 & \it$<$0.6 \\ \hline
TIMET HN3469-T & ASTM Grade 1 Slab  & \it$<$1.6 & \it$<$0.09 &- & 0.28 & 0.23 & \it$<$0.5 \\ \hline
TIMET HN3469-M & ASTM Grade 1 Slab  & 2.8 & \it$<$0.10 & -&\it$<$0.20 &0.25 & \it$<$0.7 \\ \hline
PTG Sheet (1) & ASTM Grade 1 Sheet  & 47 & 2.8 & -& 2.0 & 2.8 & \it$<$1.9 \\ \hline
PTG Sheet (2) &ASTM Grade 2 Sheet & \it$<$9.9 & 3.7 & - & \it$<$0.81 & 2.4 & \it$<$2.2 \\ \hline
Bolts & Bolts   & 1300 & \it$<$6.2 & - & 160 & 160 & \it$<$37 \\ \hline
Nuts/Washers & Nuts/Washers  & 520 & \it$<$8.6 & - & \it$<$12  & 81 & \it$<$62 \\ \hline
Honeywell Sheet & ASTM Grade 1 Sheet  & 3.7 & 4.7 & - & 1.5 & 1.6 & \it$<$1.5 \\ \hline
VSMPO Disc (10\% scrap) & ASTM Grade 1 Metal & 62 & \it$<$6.2 & - & \it$<$4.1 & \it$<$4.1 & \it$<$31 \\ \hline
VSMPO Sponge & ASTM Grade 1 Sponge & 17 & 12 & - & \it$<$4.1 & \it$<$4.1 & \it$<$6.2 \\ \hline

\end{tabular}
\end{table*}

\begin{table}[h]
\centering
\footnotesize
\caption{ \small ICP-MS assay results of TIMET HN3469-M and TIMET Sheet~(4) titanium samples. Also shown for comparison are the $^{238}$U$_e$ and $^{232}$Th$_e$ (1 $\sigma$ upper limits) results from $\gamma$-screening where uncertainties are statistical only and a systematic uncertainty of up to 10\% is estimated by comparison of cross-calibration assays. \label{tab:icp-ms}}
\begin{tabular}{|c|c|c|c|} \hline
& \multirow{2}{*}{\bf{Sample}} & \bf{$^{238}$U$_e$}  & \bf{$^{232}$Th$_e$} \\ 
& & \bf{(mBq/kg)}  & \bf{(mBq/kg)} \\ \hline
\textbf{TIMET} & UCL \#1 & $2.23\pm0.15$ &  $0.10\pm0.08$  \\
\bf{HN3469-M}& UCL \#2  & $2.01\pm0.43$ & $0.15\pm0.07$ \\ 
& HPGe  & $2.80\pm0.15$ & \textit{$<$0.20} \\
& &  &\\ 
& TiLarge Inner & $9.9 \pm 0.2$ & $0.12 \pm 0.01$  \\
 \textbf{TIMET}& TiLarge Etch & $9.9 \pm 0.8$ & $0.7 \pm 0.2$\\
\bf{Sheet (4)}& TiSmall & $9.8 \pm 0.2$  & $0.086 \pm 0.003$ \\
& HPGe & $8.0 \pm 2.0$  & \textit{$<$0.12} \\ \hline
\end{tabular}
\end{table}

\begin{table*}[t]
\footnotesize
\centering
\caption{ \small \label{tab:ss-samples} Radioassay results from the 13 stainless steel samples directly from NIRONIT, 3 from GERDA and 3 from NEXT. All limits are 1$\sigma$ upper limits  and systematic uncertainties are estimated to be up to 10\% by comparing the cross-calibration results of the assay of standard sources.
NIRONIT samples that were excluded following pre-screening due to their high $^{60}$Co content  were not assayed for early-chain contents. Samples (2--6) were selected for additional assay underground at SURF using \textsc{Maeve} primarily upon their lower $^{60}$Co content. Published results from the GERDA and NEXT collaborations are included for comparison.}
\vspace{-5pt}
\begin{tabular}{|l|c|c|c|c|c|c|} \hline
\multirow{2}{*}{\bf{Name}} & \multicolumn{2}{|c|}{\bf{$^{238}$U (mBq/kg)}} & \multicolumn{2}{|c|}{\bf{$^{232}$Th (mBq/kg)}}& \bf{$^{60}$Co} & \bf{$^{40}$K} \\
& \parbox{1.25cm}{\centering \bf early} &  \parbox{1.25cm}{\centering \bf late} & \parbox{1.25cm}{\centering \bf early} &  \parbox{1.25cm}{\centering \bf late} & \bf{(mBq/kg)} & \bf{(mBq/kg)} \\ \hline 
NIRONIT (1) & 7.3 & 0.35 & 1.1 & 4.0 & 14.5 & 0.53 \\ \hline
NIRONIT (2) & 1.2 & 0.27 & 0.12 & 0.49 & 1.6 & \it$<$0.4 \\ \hline
NIRONIT (3) & \it$<$1 & 0.54 & 0.49 & 1.1 & 1.7 & \it$<$0.59 \\ \hline
NIRONIT (4) & 1.4 & 0.5 & 0.5 & 0.32 & 2.6 & \it$<$0.5 \\ \hline
NIRONIT (5) & 1.1 & 0.38 & 0.81 & 0.73 & 5.6 & \it$<$0.46 \\ \hline
NIRONIT (6) & 0.5 & 1.9 & 1.7 & 1.5 & 4.5 & \it$<$0.5 \\ \hline
NIRONIT (7) & - & 1.1 & - & 4.1 & 8.2 & \it$<$3.0 \\  \hline
NIRONIT (8) & - & \it$<$0.6 & - & \it$<$0.8 & 7.4 & \it$<$3 \\  \hline
NIRONIT (9) & - & \it$<$0.6 & - & \it$<$0.9 & 6.5 & \it$<$3 \\ \hline
NIRONIT (10) & - & 4 & - & 2.2 & 26 & \it$<$4 \\ \hline
NIRONIT (11) & - & \it$<$0.6 & - & 4.8 & 32 & \it$<$2 \\ \hline
NIRONIT (12) & - & \it$<$0.8 & - & 2.1 & 32 & 5 \\ \hline
NIRONIT (13) & - & \it$<$1.4 & - & \it$<$1.5 & 335 & \it$<$4 \\ \hline
GERDA  D6 & \multicolumn{2}{|c|}{ $<$5 } & \multicolumn{2}{|c|}{\it$<$0.4} & - & \it$<$0.002 \\ published \cite{maneschg:2008zz} & \multicolumn{2}{|c|}{\it$<$0.6} & \multicolumn{2}{|c|}{\it$<$1.4} & $16.8 \pm 2.4$& \it$<$1.8 \\ \hline
GERDA G1 & \multicolumn{2}{|c|}{ $<$5 } & \multicolumn{2}{|c|}{\it$<$0.4} & - & \it$<$0.003 \\ published \cite{maneschg:2008zz} & \multicolumn{2}{|c|}{\it$<$1.3} & \multicolumn{2}{|c|}{\it$<$2.6} & $45.5\pm2.1$ & \it$<$2.8 \\ \hline
GERDA G2 & \multicolumn{2}{|c|}{ $<$5 } & \multicolumn{2}{|c|}{\it$<$0.4} & - & \it$<$0.003 \\ 
published \cite{maneschg:2008zz} & \multicolumn{2}{|c|}{\it$<$0.86} & \multicolumn{2}{|c|}{\it$<$0.24} & $14.0\pm0.1$ & $<0.93$ \\ \hline
NEXT 10 mm & \multicolumn{2}{|c|}{ 7.46 } & \multicolumn{2}{|c|}{\it$<$0.24} & - &\it$<$0.63 \\
 published \cite{alvarez:2013rp} & \multicolumn{2}{|c|}{ \it$<$21 } & \multicolumn{2}{|c|}{\it$<$0.59} &$2.8\pm0.2$ & \it$<$0.96 \\ \hline
NEXT 15 mm & \multicolumn{2}{|c|}{ $12.4$ } & \multicolumn{2}{|c|}{\it$<$0.24} & - & \it$<$0.63 \\
 published \cite{alvarez:2013rp} & \multicolumn{2}{|c|}{ \it$<$25 } & \multicolumn{2}{|c|}{\it$<$0.69} &$4.4\pm0.3$ & \it$<$1.0 \\ \hline
NEXT 50 mm & \multicolumn{2}{|c|}{ $12.4$ } & \multicolumn{2}{|c|}{\it$<$0.24} & - & \it$<$0.63 \\ 
 published \cite{alvarez:2013rp} & \multicolumn{2}{|c|}{ $67 \pm 22$ } & \multicolumn{2}{|c|}{$2.1 \pm 0.4$} &$4.2\pm0.3$ & \it$<$2.5 \\ \hline
\end{tabular}
\end{table*}

ICP-MS assays have been performed to directly measure the $^{238}$U and $^{232}$Th content of the most radio-pure samples as identified through $\gamma$-screening measurements:  TIMET HN3469 and TIMET Sheet (4), both ASTM Grade 1 titanium with 0\% scrap, and produced with EBCH re-melting. ICP-MS can achieve higher sensitivity for some elements in particular matrixes, utilizes a much smaller sample mass, and requires a shorter measurement time than in $\gamma$-screening. However, ICP-MS can suffer systematic errors that are larger than those in $\gamma$-screening measurements, due to contamination of the smaller sample mass, and variation due to the sampling of a small fraction of a large mass. Two pieces of TIMET HN3469-M were assayed with ICP-MS at the UCL facility. The samples were sonicated in high purity IPA before rinsing in de-ionised water, followed by an etch in 1:1 HF:HNO$_3$ acid, further rinsing, and drying in an ISO class 10 LFU. The prepared samples, with masses 128.8 g (UCL \#1) and 202.3 (UCL \#2), were digested in a mixture of de-ionised water, HF, and HNO$_3$, with acids doubly distilled. The digestion was performed in a closed microwave oven, with samples spiked with $^{233}$U and $^{230}$Th tracers assayed to monitor and correct for uranium and thorium recovery, and separate sample blanks were also assayed for background subtraction.. The results of the assays are shown in Table II, with detections of $^{238}$U at $2.23\pm0.37$~mBq/kg and $^{232}$Th at $0.10\pm0.08$~mBq/kg in sample UCL \#1, and  $^{238}$U at $2.01\pm0.43$~mBq/kg and $^{232}$Th at $0.15\pm0.07$~mBq/kg in sample UCL \#2. An additional 15\% systematic uncertainty is estimated based on ICP-MS assays of calibration standards, such as IAEA-385~\cite{IAEA385}, used in the HPGe cross-calibration campaign. The results are consistent with the HPGe assays of TIMET HN3469-M, also shown in Table II for comparison, at $2.80\pm0.15$~mBq/kg for $^{238}$U$_e$ and  at $<$0.20~mBq/kg for $^{232}$Th$_e$ indicating the successful control of systematic effects in the ICP-MS process.  Similarly, two samples of TIMET Sheet (4) were assayed at PNNL: a large piece (TiLarge) separated into 3 pieces of 1 g each measured separately, and a small rectangular bar (TiSmall) measured once. A surface etch was performed on the TiLarge pieces to remove any surface contamination remaining after cleaning the samples. The etched material (TiLarge Etch) was retained and assayed separately to the bulk material (TiLarge Inner). The results of the assays are also presented in Table II. Uncertainties reported for TiLarge use the standard deviation of the three samples as well as instrumental precision, whilst TiSmall uses instrumental precision only. $^{238}$U results are consistent between samples, whilst $^{232}$Th results show discrepancy particularly from the etched material, likely due to surface contamination from processing. The results from the the ICP-MS are generally in agreement with the LZ $\gamma$-screening measurements of $8\pm2$ mBq/kg of $^{238}$U and $<$0.12 mBq/kg of $^{232}$Th again highlighting good control of systematic effects in the ICP-MS process.

In parallel with the titanium R\&D, we maintained a programme to secure samples of stainless steel as a viable alternative for the LZ cryostat to  mitigate risk of being unable to find a source of sufficiently radiopure titanium, or an inability to manufacture a sufficiently large quantity. Stainless steel samples were sourced from NIRONIT~\cite{NIRONIT:2015}, suppliers for the GERDA~\cite{maneschg:2008zz}, NEXT~\cite{alvarez:2013rp} and XENON100~\citep{aprile:2011vb} experiments. We also re-measured a selection of stainless steel samples previously measured by the GERDA and NEXT experiments for independent assay, finding activity levels consistent with theirs~\cite{maneschg:2008zz, alvarez:2013rp}. The radioassay results of the 22 stainless steel samples are presented in Table~\ref{tab:ss-samples}. NIRONIT samples were sourced from ThyssenKrupp (samples 2--6, 9 and 13) and Aperam (1, 7, 8, 11, 12).

\subsection{Material Selection}
Following the assays of all 22 Ti and 22 stainless steel samples, the highest and most reproducible radiopurity is found in ASTM Grade 1 titanium that contained 0\% scrap and had been refined using EBCH technology. Of these, the lowest radioactivity is observed in material from TIMET Heat Number (HN) 3469, a single 15,000~kg slab of titanium, produced by TIMET at its Morgantown (Pennsylvania) mill. Results from a chemical analysis of this sample to measure  concentrations of iron, carbon, oxygen and nitrogen are  shown in Table~\ref{tab:TIMET}.

\begin{table}[h]
\centering
\footnotesize
\caption{ \small Chemical analysis of two samples from TIMET HN3469-T, showing \% weight of iron, carbon, oxygen and nitrogen in the titanium \cite{TIMET:2016}. \label{tab:TIMET}}
\begin{tabular}{|c|c|c|c|c|} \hline
\bf Sample & \bf Fe & \bf C & \bf O & \bf N \\ \hline
1 & 0.01 & 0.001 & 0.04 & 0.002 \\
2 & 0.02 & 0.006 & 0.02 & 0.002 \\ \hline
\end{tabular}
\end{table}

A sample from TIMET HN3469, denoted HN3469-T, was screened at the Berkeley Low Background Facility using \textsc{Maeve} in May of 2015. This consisted of 10.1~kg of plates selected from the top portion of the single slab. A second sample taken from the middle (HN3469-M), was acquired and assayed in September of 2015 to confirm the uniform distribution of contamination. The radioactivities of both samples of this titanium stock were found to be consistent.

In both samples, each counted for approximately three weeks, the early uranium chain was non-detectable or barely detectable within the limited abilities of HPGe detectors to assay this portion of the decay chain via $\gamma$-ray spectroscopy. The late portion of the chain (at $^{226}$Ra and below), however, is quite accessible via $\gamma$-ray spectroscopy due to both the branching ratio and detection efficiency for $\gamma$-rays emitted during the decay of constituent isotopes, and registered no detectable activity above background down to a few ppt. The $^{238}$U$_l$ value is based upon the 609~keV peak from $^{214}$Bi and is consistent when compared to upper limits from other useful peaks in the late uranium chain, such as from $^{214}$Pb (295, 352~keV) and $^{214}$Bi (1764~keV). Both samples had detectable levels of the thorium series in both the early and late portions of the chain, consistent with secular equilibrium in both samples. The $^{232}$Th$_l$ measurement is based upon the 238~keV $\gamma$-ray from $^{212}$Pb, which is the strongest peak given the product of the detection efficiency of that $\gamma$-ray line and its branching ratio in the thorium chain. The assays are summarized in Table~\ref{tab:titaniumHN3469}.  

In terms of cosmogenic activation there were several isotopes of scandium present, most of which are the result of cosmic ray-induced reactions with the five stable isotopes of titanium. Detected in the sample was $^{46}$Sc (889, 1121~keV, $T_{1/2}$=84 days); as well as small amounts of  $^{47}$Sc (159~keV, $T_{1/2}$=3 days),  $^{48}$Sc (984, 1038, 1312~keV, $T_{1/2}$=44 hours), and $^{44,44m}$Sc (271, 1157~keV, $T_{1/2}$=59 hours and 4 hours, the metastable state being the longer-lived). Screening results for $^{46}$Sc are also listed in Table~\ref{tab:titaniumHN3469}, whilst the $^{47}$Sc, $^{48}$Sc, and $^{44,44m}$Sc activities are not listed as their short half lives mean they essentially disappear over the course of the measurement. The reported value for $^{46}$Sc was corrected to the start of counting for each of the samples. The cosmogenic production of $^{46}$Sc can be mitigated by moving components underground as soon as possible after manufacture.   

\begin{table}[h]
\caption{ \small \label{tab:titaniumHN3469} Results from $\gamma$-ray spectroscopy (in mBq/kg) of TIMET titanium sample HN3469 for both top and middle samples. All limits are 1$\sigma$ upper limits and uncertainties are statistical only. A systematic uncertainty of up to 10\% has to be additionally assumed.  Results were obtained with the \textsc{Maeve} detector and confirmed with ICP-MS and the \textsc{Chaloner} detector.}
\begin{center}
\footnotesize
\begin{tabular}{|l|c|c|}  
 \hline
 & \textbf{Top} & \textbf{Middle} \\
 \hline
\bf Date & May 2015 & Sep. 2015 \\
\bf Sample mass & 10.07~kg & 8.58~kg \\
\bf Livetime & 23.9~days & 20.8~days \\ \hline
\multicolumn{3}{|c|}{\bf{Activities (mBq/kg)}} \\ \hline
\bf $^{238}$U$_{e}$  & \it$<$1.6 & $2.80\pm 0.15$ \\
\bf $^{238}$U$_{l}$  & \it$<$0.09 & \it$<$0.10 \\
\bf $^{232}$Th$_{e}$  &$ 0.28\pm0.03$ & \it$<$0.20 \\
\bf $^{232}$Th$_{l}$  & $0.23\pm0.02$ & $0.25\pm0.02$ \\
\bf $^{40}$K   & \it$<$0.54 & \it$<$0.68 \\
\bf $^{60}$Co  & \it$<$0.02 & \it$<$0.03 \\
\bf $^{46}$Sc  & $2.0\pm 0.1$ & $2.7\pm0.1$ \\
\hline
\end{tabular} 
\vspace{-20pt}
\end{center}
\end{table}
The activities reported here for TIMET HN3469 are among the lowest worldwide known to-date for titanium; the radiopurity is comparable to the LUX titanium (U$_{l}$$<$0.25 mBq/kg, Th$_{l}$$<$0.2 mBq/kg \cite{akerib:2011rr}), making the campaign a success. Titanium from the HN3469 slab has been procured for the 2,292~kg LZ cryostat and cryostat support structures, as well as other internal structures amounting to $\sim$364~kg. The measured radioactivities in the cryostat materials used in other dark matter and $0\nu\beta\beta$ experiments are shown for comparison in Table~\ref{tab:comparison}. The cryostat material with the lowest radiopurity and most sensitive measurements is the copper used in the Majorana Demonstrator (MJD) $0\nu\beta\beta$ experiment, which is electroformed underground to ensure extremely low levels of contamination.

\begin{table*}[t]
\centering
\footnotesize
\caption{\small Radioactivities of cryostat material for a selection of dark matter and 0$\nu\beta\beta$ experiments. Late chain {$^{238}$U} \& {$^{232}$Th} activities are quoted due to the sparsity of data on early chain activities. Activities for  {$^{60}$Co}, {$^{40}$K}, and {$^{46}$Sc} are included where results are publicly available. Results from XENON100 and GERDA were reported as total activity and have been scaled by the given mass for comparison.  \label{tab:comparison}}
\begin{tabular}{|c|c|c|c|c|c|c|c|c|} \hline
\multirow{2}{*}{\bf{Experiment}} & \multirow{2}{*}{\bf{Type}} & \bf Mass & \bf{$^{238}$U} & \bf{$^{232}$Th} & \bf{$^{60}$Co} & \bf{$^{40}$K} & \bf{$^{46}$Sc} \\ 
& & \bf(kg) &   \bf{(mBq/kg)}& {\centering \bf{(mBq/kg)} } &\bf{(mBq/kg)} & \bf{(mBq/kg)} &  \bf{(mBq/kg)} \\\hline

 DarkSide50~\cite{DARKSIDE50:2015} & SS & 175 & \it$<$1 &\it$<$1 & $13.1 \pm 1 $ & - & -  \\\
 XENON100~\cite{aprile:2011vb} & SS & 74 & \it$<$1.8 & \it$<$0.03 & $5.4\pm0.5$  & \it$<$9  & - \\ 
 XENON1T~\cite{Aprile:2015uzo} & SS & 870 & $2.4\pm0.7$   & $0.21\pm0.06$  & \it$<$0.36  & $9.7\pm0.8$ & 2.7 \\ 
ZEPLIN-III~\cite{araujo:2011as} & Cu & 400 &\it$<$6.22 &\it$<$2.03   & - & $<$\it0.32& - \\ 
EXO-200~\cite{EXO:2012} & Cu & 5,901 & \it$<$0.01& \it$<$0.003 & - & \it$<$0.01 & - \\
\multirow{2}{*}{GERDA~\cite{Agostini:2013tek}} & Cu & 16,000 & $0.017\pm0.005$  &$0.014\pm0.005$ &$0.018\pm0.005$ &$<$0.049 & - \\
& SS & 25,000 & \it $<$1.2 & \it $<$1.2 &  19 & \it  $<$2.9 & - \\
MJD~\cite{Abgrall:2016act} & Cu & 1,297 & \it$<$0.0003& \it$<$0.0003 & - & - & - \\
LUX~\cite{akerib:2011rr} & Ti & 230 & \it$<$0.25 & \it$<$0.2 & - & \it$<$1.2& $2.5$\\ 
\multirow{2}{*}{LZ (this work)} & \multirow{2}{*}{Ti} & \multirow{2}{*}{1,827} & U$_e$ $<$\it1.6 & Th$_e$: $0.28\pm0.03$   & \multirow{2}{*}{$<$\it0.02} & \multirow{2}{*}{$<$\it0.54} & \multirow{2}{*}{$2.0\pm0.1$}\\ 
 & & &U$_l$ $<$\it0.09 & Th$_l$: $0.23\pm0.02$& & &  \\
\hline

\end{tabular}
\end{table*}

\section{Cryostat Background}
\label{sec:LZSim}
\subsection{Overview of Backgrounds}
Backgrounds from the cryostat may be categorised into two classes: electron recoils (ER) and nuclear recoils (NR). ER events can be discriminated against in a LXe TPC to greater than 99.5\% efficiency, while NR events may not as they are essentially signal-like. Neutron-induced NR are indistinguishable from WIMPs if only a single elastic scatter occurs within the LZ fiducial volume, unaccompanied by any other signal in other active veto volumes of the experiment. Leading LZ background sources include astrophysical neutrinos, intrinsic radioactivity in the Xe, and emission from every component in the experiment. 

Radon emanation from the cryostat materials contributes an additional ER background, due to the `naked' beta emission from $^{214}$Pb in the $^{222}$Rn sub-chain. Background may also be generated from residual dust on the detector surfaces and plate-out of radon daughters, by emission of $\gamma$-rays from the radioactivity in the dust and neutrons from ($\alpha$,$n$) reactions, as well as  further radon emanation into the LXe from the dust, specifically from the inside of the inner vessel. Plate-out of radon progeny onto the surfaces of the cryostat following final cleaning post-manufacture must also be considered, especially the long-lived isotope $^{210}$Pb. Daughters of $^{210}$Pb, specifically $^{210}$Po, can induce neutron emission through ($\alpha$,$n$) reactions anywhere on the titanium surfaces, despite the relatively low $\alpha$ energy, generating additional nuclear recoil background. Furthermore, the presence of a reflective PTFE liner inside the inner vessel must also be considered; neutrons can also be induced here by $\alpha$-particles from the cryostat, a process enhanced by the relatively high $(\alpha,n)$ cross section of the fluorine present in PTFE. 

\subsection{Background Modeling}
Monte Carlo simulations have been performed to assess the contribution from all expected background sources, particularly those that generate single vertex interactions as expected from WIMP elastic scattering. These are conducted using LZSim, a simulation package constructed to model the experiment and inform the design, determine optimal performance parameters, and define tolerable rates from background sources. Developed using the \textsc{Geant4} toolkit~\cite{Agostinelli:2002hh}, the framework inherits from the LUX model~\cite{Akerib:2011ec}, and includes all parts of the experiment.

\begin{figure}[h]
\centering
\begin{minipage}[tl]{0.23\textwidth}
\includegraphics[height=155pt, trim = 0 40 0 40]{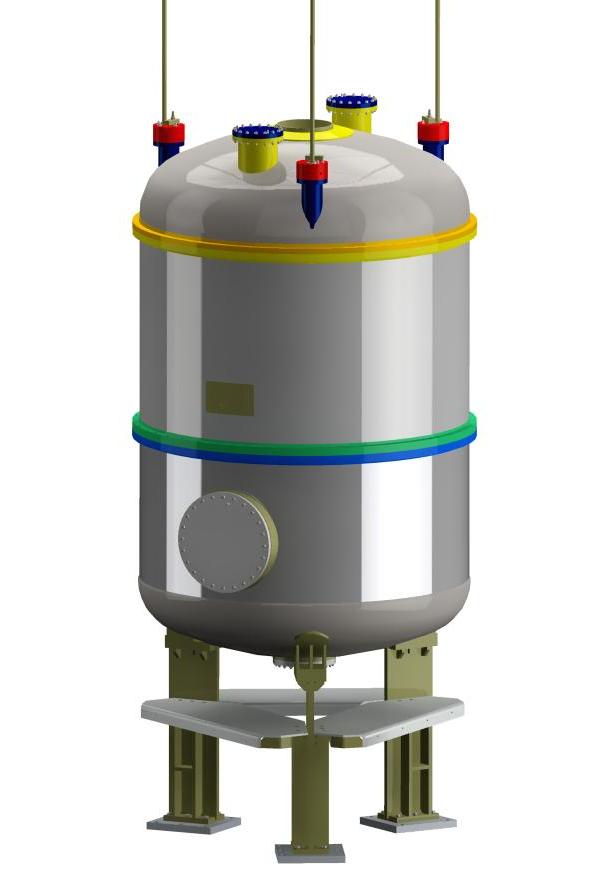}
\subcaption{OCV and CS}
\end{minipage}
\begin{minipage}[tr]{0.23\textwidth}
\includegraphics[height=155pt, trim = 20 0 40 0]{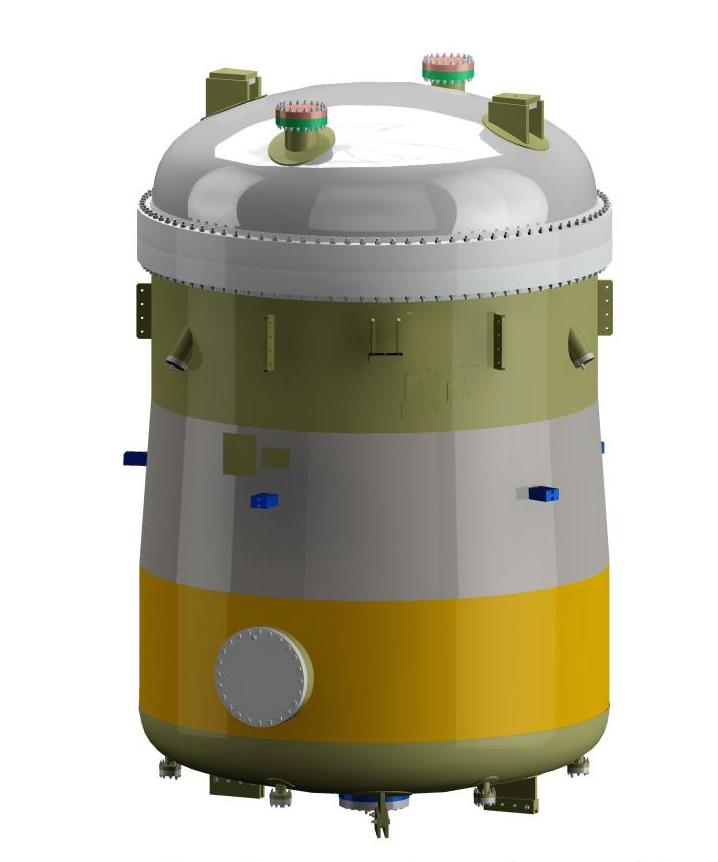}
\subcaption{ICV}
\end{minipage}
\caption{\small CAD models of the cryostat vessels and support. The OCV connects to three support legs mounted on baseplates and the ICV will be placed inside the OCV in vacuum. The full cryostat assembly stands 4.1~m tall.  \label{fig:cryocad}}
\end{figure}

The LZ cryostat has been designed to stringent requirements that factor science goals, safety, containment, installation, and operation, and complies with ASME Boiler Pressure and Vessel Code (BPVC) VIII Div.~1~\cite{ASMEweld:2015}. It will contain a total of 10 tonnes of LXe at a minimum temperature of $-110$\textdegree C and be placed inside a water tank for shielding purposes. There are three primary parts---the inner cryostat vessel (ICV), the outer cryostat vessel (OCV) and the cryostat support (CS). Both the OCV and ICV contain a series of ports for cabling, liquid exchange and calibration sources.  A CAD model of the cryostat is shown in Figure~\ref{fig:cryocad}.

\subsection{Radioactivity Simulations}
Simulations have been conducted by positioning radioactive nuclides in all the cryostat materials including the titanium, seals, liners, insulation, and bolts. For ER simulations, $^{60}$Co, $^{40}$K and the U and Th chain decays are simulated to produce $\gamma$-rays with the correct energies and branching ratios. For the U and Th early and late sub-chains, activities are assigned as determined by assays. Additionally,  radon decays inside the LXe are simulated, under the assumption that all radon released from any material in contact with Xe mixes with the active mass.

For NR simulations, a neutron ($\alpha,n$) energy spectrum for each material was obtained from SOURCES4A \cite{Wilson:2009}. Single neutrons were emitted with energies sampled from the relevant spectrum, depending on their source material. The SOURCES4A code has been modified to extend the energies of $\alpha$-particles up to 10 MeV from the original upper energy cut at 6.5 MeV \cite{carson:2004cb}, and to improve and extend the ($\alpha,n$) cross-section library for a large number of materials~\cite{lemrani:2006dq,tomasello:2008ri,tomasello:2010zz}, with newly added cross-sections calculated using the EMPIRE-2.19 code \cite{herman:2007a}. In calculations of neutron yields we use the thick target approximation, valid for material thicknesses significantly exceeding the range of $\alpha$-particles, which are 10--30~$\mu$m in titanium. We find $<$0.1\% of $\alpha$-decays in the titanium are able to reach the neighbouring PTFE and induce a neutron, and so we consider this rate as negligible.

In principle, $^{238}$U spontaneous fission can contribute to neutron yields significantly in materials where ($\alpha,n$) rates are low. However, the near-simultaneous emission of up to 5 neutrons and up to 20 $\gamma$-rays results in highly efficient event rejection, reducing the contribution from spontaneous fission to negligible levels, and so these are removed from the background model~\cite{shaw:2016}. In our simulations, this reduces the neutron yield in titanium by 35\%. 

The simulation output is analysed to examine events with energy depositions in the LXe active target. In particular, for an event to constitute a background count, the energy deposited in the active LXe must be within the WIMP search region of interest (6--30 keV recoil energy) and the particle must be observed to have scattered only once within the active LXe volume. Furthermore, LZ will have a veto detector external to the cryostat, and this may not register an energy deposition within an event time window.

\begin{figure*}[t]
\centering
\begin{minipage}{0.48\textwidth}
\includegraphics[width=\linewidth,trim = 0 180 0 170]{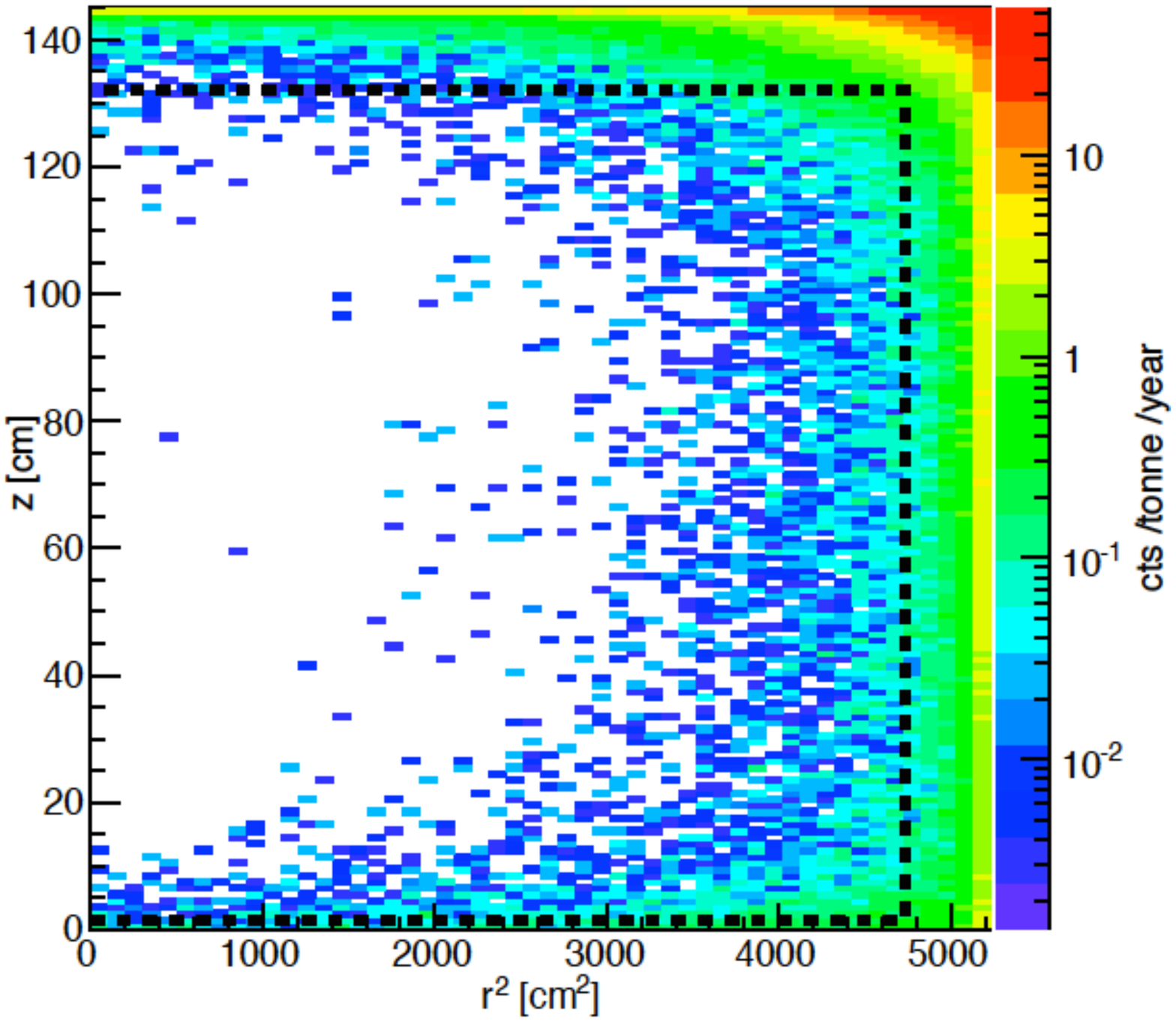}
\subcaption{Single scatter energy deposits.}
\end{minipage}
\begin{minipage}{0.48\textwidth}
\includegraphics[width=\linewidth,trim = 0 180 0 170]{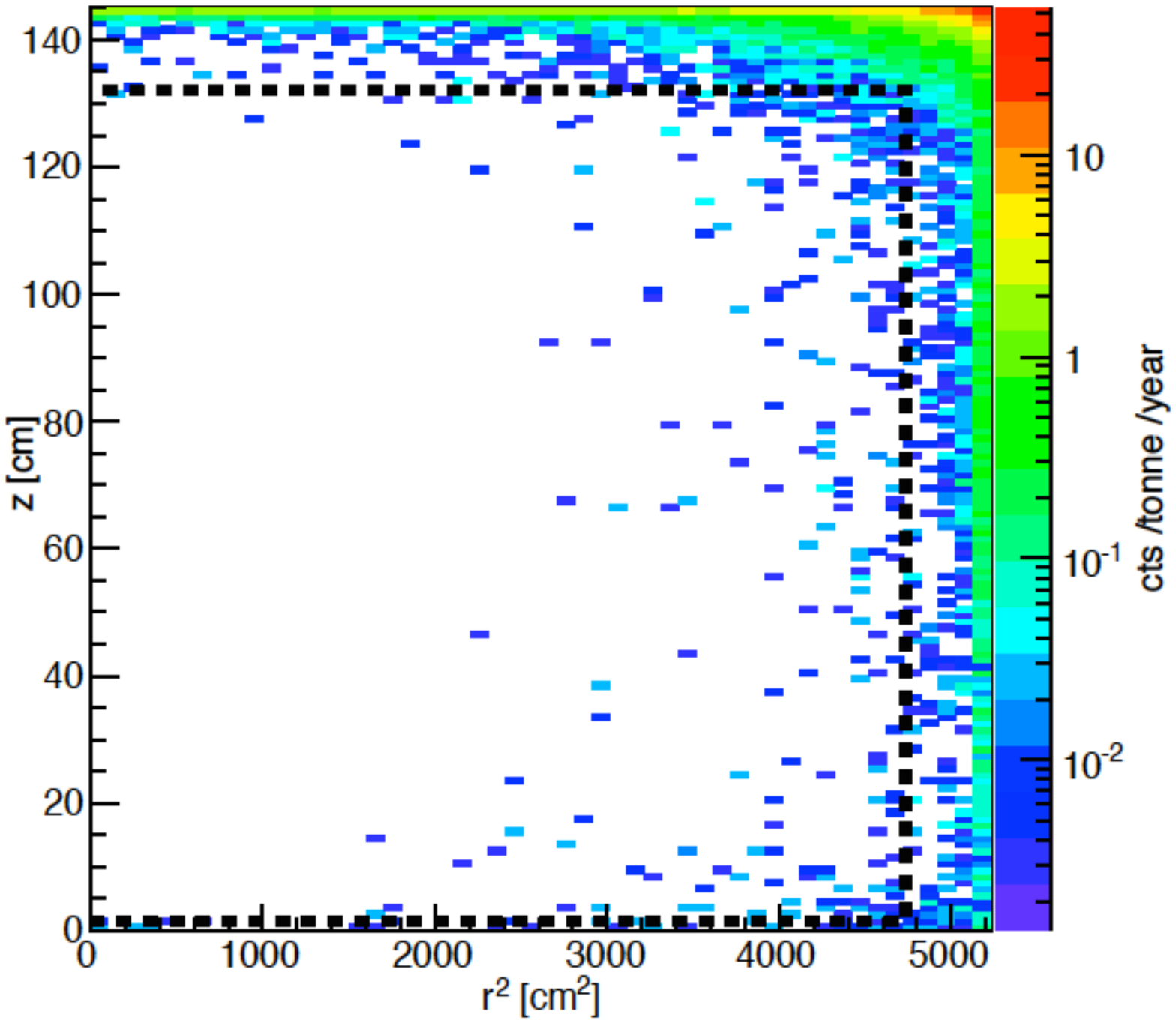}
\subcaption{Single scatter energy deposits after vetoing.}
\end{minipage}
\caption{\label{fig:cryotemp} \small Expected background from the cryostat in LZ, showing the spatial distribution in $r^2$ and $z$ within the active volume, and the fiducial volume (black dashed) that contains 5.6 tonnes of LXe. The colour scales are in counts/tonne/year. The effect of vetoing  is highlighted through the reduction of events that satisfy single scatter selection in the energy range of interest for a WIMP search (left) by a factor $\sim 10$ when vetoes are implemented  (right).}
\end{figure*}

Table~\ref{tab:backgrounds} summarises the materials used in the cryostat simulations, their masses, their measured activities, the neutron emission rate as determined with SOURCES4A for the given activities and mass, and the contribution to ER and NR background in LZ following the selection criteria.
Figure~\ref{fig:cryotemp} shows the spatial distribution of events generated by the cryostat radioactivity in the 7 tonne active volume and energy range of interest, showing the reduction as single scatter selection and vetoing is implemented.  
\begin{table*}[t]
\centering
\footnotesize
\caption{ \small Table of expected backgrounds from all cryostat components, in an exposure of 5,600 tonne-days, before discrimination and NR acceptance is applied. The stainless steel is used for fasteners and calibration tubes, and the PEEK for seismic limiters.  The totals show the mass weighed activities. ER and NR counts are rounded to two decimal places, and statistical uncertainties are given for the totals.  \label{tab:backgrounds}}
\begin{tabular}{|l|c|c|c|c|c|c|c|c|c|} \hline
\multirow{2}{*}{\bf{Name}} & \bf{Mass}  & \multicolumn{2}{|c|}{\bf{$^{238}$U (mBq/kg)}} & \multicolumn{2}{|c|}{\bf{$^{232}$Th (mBq/kg)}}& \bf{$^{60}$Co} & \bf{$^{40}$K}  & \bf{ER} & \bf{NR} \\
 & \bf (kg) &  \parbox{1 cm}{\centering \bf early} & \parbox{1 cm}{ \centering \bf late} &  \parbox{1 cm}{ \centering \bf early} & \parbox{1 cm}{ \centering \bf  late} & \bf (mBq/kg) & \bf (mBq/kg) & \bf(cts)  & \bf(cts) \\ \hline 
 Titanium  & 2,292 & 1.59 & 0.11 & 0.29 & 0.25 & 0.00 & 0.54  & 0.51 & $1.27\times10^{-2}$\\
 S. Steel & 110.5 & 1.429 & 0.27 & 0.33 & 0.49 & 1.60 & 0.40 & 0.09 & $3.01\times10^{-4}$ \\
 PEEK & 1.167 & 17.00 & 16.60 &  16.10& 8.50 & 0.52 & 40.80  & 0.02 &$1.42\times10^{-4}$ \\ \hline
 \textbf{Total} &  2,404 & 1.59 & 0.11 & 0.29 & 0.25 & 0.07 & 0.56  & \textbf{0.63$\pm$0.06} & \textbf{0.013$\pm$0.001 }\\ \hline
\end{tabular}
\end{table*}
\subsection{Total Cryostat Background}
\begin{figure*}[t]
\centering
\includegraphics[width=1\textwidth, trim = 0 0 0 0]{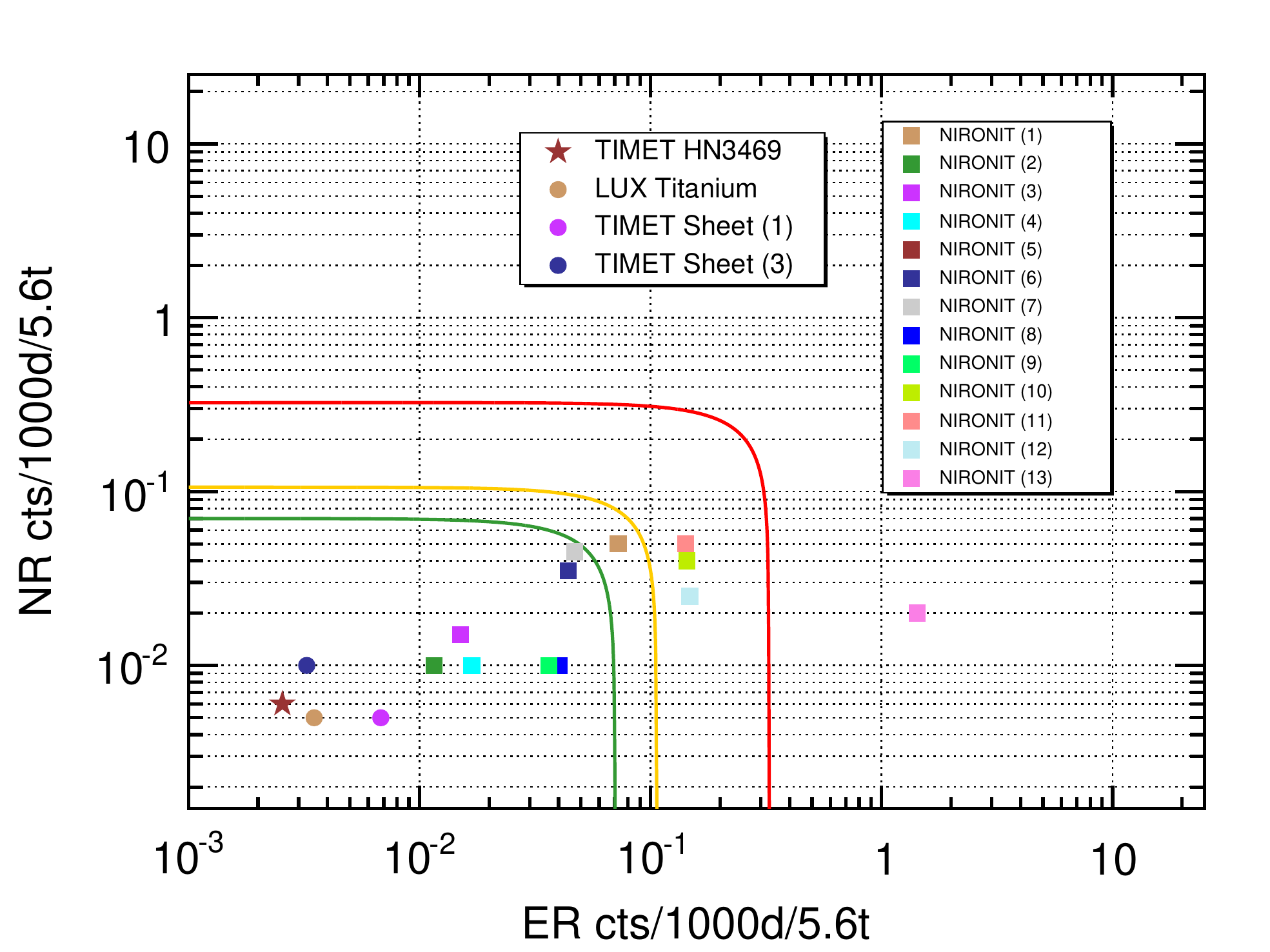}
\caption{\small Expected background counts in ER and NR, within the fiducial volume after all vetoes are applied, showing the TIMET sample chosen for LZ, the screened stainless steel samples as well as the titanium used for the LUX cryostat. An NR acceptance of 50\% and an ER discrimination factor of 99.5\% are applied. The red curve corresponds to the sum of 10\% of the \textit{pp} solar neutrinos and 0.2 NR events, the yellow is the sum of 5\% of the \textit{pp} solar neutrino ER background and 0.05 NR events, and the green line is the sum of 3.3\% of the \textit{pp} solar neutrino ER background and 0.03 NR events. This green line was the requirement for the LZ cryostat, and the titanium chosen (indicated by the star) is well below this requirement. \label{fig:TiERNR}}
\end{figure*}

The total expected background in LZ's 5.6 tonne fiducial mass in a 1,000-day exposure (after the application of the region of interest, single scatter and veto cuts described above) from the intrinsic radioactivity within the titanium of the  cryostat is  0.0127 NR and 0.51 ER counts. Discrimination allows 99.5\% rejection of ERs, reducing the cryostat contribution of WIMP-like events from ERs to 0.003 events.   
The NR background is also reduced to 0.006 counts under the standard NR acceptance of $\sim$50\% typically assumed in LXe TPC dark matter experiments. Figure~\ref {fig:TiERNR} shows this background contribution together with that expected if a selection of alternative titanium samples were to be used based on our radioassay results. Also included is a hypothetical point based on radioactivity equal to the LUX cryostat material. The ER and NR background expected if any of the stainless steel samples were to be used for the LZ cryostat are also included. Contours on the figure indicate goals on background contribution from the cryostat material. The red contour represents 0.2 NR counts and an ER rate equivalent to 10\% of the background expected from \textit{pp} solar neutrino scattering in the active mass---one of the largest contributions to the experiment. This marks the upper limit set on background from intrinsic radioactivity fixed in all materials in the LZ experiment. The yellow contour is 0.05 NR and 5\% of the \textit{pp} solar neutrino rate. Given the cryostat mass and proximity to the LXe, it is allocated up to a third of the maximum rate, 0.03 NR and 3.3\% of the \textit{pp} solar neutrino rate, denoted by the green contour. The TIMET samples produced with EBCH and 0\% scrap and the LUX titanium each satisfy this goal. Although some of the stainless steel samples assayed would be acceptable, most either fail the requirement or are very close to failing, and all exceed rates from titanium. Furthermore, the large sample-to-sample variation indicates poorer reproducibility under standard manufacturing and processing. This presents increased risk of differences in  radioactivities between assayed and procured material. The use of a stainless steel cryostat also reduces the efficiency of ER background vetoing, with $\gamma$-rays stopped in the vessels rather than penetrating to veto systems. 

The ER and NR contributions from the intrinsic radioactivity fixed within all of the cryostat components is presented in Table~\ref{tab:backgrounds}, after all cuts are applied. The combined contribution is $0.013\pm0.001$ NR and $0.63\pm0.07$ ER counts, reduced to $0.0032\pm0.0004$ WIMP like ER counts and $0.0065\pm0.0005$ NR counts after 50\% NR acceptance and 99.5\% ER discrimination.

Estimates of radon emanation from titanium have been made using emanation rates from steel in literature and conservative models of the reduction expected at cryogenic temperatures~\cite{Zuzel:2005}. The total estimated activity from the titanium cryostat from  $^{222}$Rn is 0.15 mBq. This activity is combined with radon emanation from the dust on the inside of the ICV, assuming typical $^{238}$U concentrations of $\approx$10~mBq/g and a surface density of 500~ng/cm$^2$, for a conservative total of 30 ER counts. 

The contribution from intrinsic dust activity is insignificant, contributing only $1.5\times 10^{-4}$ NR counts and $7.6\times 10^{-3}$ ER counts. This was calculated assuming conservative exposure periods to air with no radon mitigation. Radon daughter plate-out, which leads to contamination of titanium surfaces with the long-lived isotope $^{210}$Pb, can induce neutron emission through ($\alpha$,$n$) reactions, but given the efficiency of neutron vetoing and the single scatter selection criteria in the fiducial volume, this process is expected to contribute only 2.8$\times 10^{-3}$ NR counts.

The combined total activity from the cryostat including fixed radioactivity, radon emanation, plate-out and dust is 0.152 ER and 0.008 NR after discrimination and acceptances. Models of dust deposition are assumed to have a systematic uncertainty of 30\%. These and the systematic uncertainty on radioassays of 10\% are combined with statistical uncertainty from simulations to give a total of $0.160\pm0.001$(stat)$\pm0.030$(sys) background counts to the WIMP search exposure of 5,600 tonne-days. The breakdown from the contributors to this total is presented in Table~\ref{tab:summary}.  

\begin{table}[h]
\centering
\footnotesize
\caption{\small Breakdown of expected counts from the cryostat, in 5.6 tonnes of LXe and 1000 days; the total is shown before and after an ER discrimination of 99.5\%, and an NR efficiency 50\%. \label{tab:summary}}
\begin{tabular}{|l|c|c|} \hline
  & \bf NR (cts) & \bf ER (cts) \\ \hline
Titanium contamination & $1.27\times10^{-2}$ & 0.51 \\
Other contamination & $4.43\times10^{-4}$ & 0.11\\
Rn emanation & 0 & 29.7\\
Rn daughter plate-out & $2.81\times10^{-3}$ & 0 \\
Dust & $1.51\times10^{-4}$ & $7.58\times10^{-3}$ \\ 
Total (before dis./acc.)  & 0.016 &  30.31 \\ \hline
\bf Total (after dis./acc.) & \bf 0.008 & \bf 0.152 \\ \hline

\end{tabular}
\end{table}
\section{Conclusion}
The LZ collaboration has completed an R\&D campaign to identify and procure radiopure titanium for construction of a double vessel cryostat. Material will also be used for construction of other internal detector components. We identified titanium made from EBCH melt with 0\% scrap, produced by TIMET, as by far the best material. The measured activities for $^{238}$U, $^{232}$Th, $^{60}$Co and $^{40}$K from the sample are significantly lower than requirements and are, alongside the LUX titanium, the lowest reported in titanium to date. 

Identification of low radioactivity titanium, together with associated manufacturing criteria required to reliably reproduce such activities, represents mitigation of a significant potential source of background for future dark matter experiments that will operate with metal cryostat vessels. 

The total contribution to LZ backgrounds in the standard LZ exposure of 5.6 tons and 1000 days from the titanium vessels alone is 0.51 ER and 0.0127 NR, before any discrimination is applied. Accounting for all ancillary components and materials, and backgrounds that may be introduced during fabrication, assembly, and installation, the cryostat is expected to contribute a background of up to 0.152 ER and 0.008 NR WIMP-like background events to the LZ science run.

\section*{Acknowledgements}

We thank Alex Chepurnov of the Scobelcyn Institute of Nuclear Physics of Moscow State University for useful discussions and insights and are grateful to TIMET and NIRONIT for their co-operation with this work. We thank the GERDA and NEXT collaborations for supplying samples, and  Mike Buhr from NIRONIT for support and securing and sending stainless steel samples.

In particular the project would like to thank Jim Grauman from TIMET for his help in getting the Ti samples and securing for us the final slab. Without his constant support we would have not succeeded in our ultra low background material search campaign. Also we would like thank him for his expertise and valuable comments on the titanium production. 
We thank Alan R. Smith of Lawrence Berkeley National Laboratory for his work on many of the measurements made during this assay campaign.

This work was partially supported by the U.S. Department of Energy (DOE) under award numbers DESC0012704,DE-SC0010010, DE-AC02-05CH11231, DE-SC0012161, DE-SC0014223, DE-FG02-
13ER42020, DE-FG02-91ER40674, DE-NA0000979, DE-SC0011702, DESC0006572, DESC0012034, DE-SC0006605, and DE-FG02-10ER46709; by the U.S. National Science Foundation (NSF) under award numbers NSF PHY-110447, NSF PHY-1506068, NSF PHY-1312561, NSF PHY-1406943 and NSF PHY-1642619; by the U.K. Science \& Technology Facilities Council under award numbers ST/K006428/1, ST/M003655/1, ST/M003981/1, ST/M003744/1, ST/M003639/1, ST/M003604/1, and ST/M003469/1; and by the Portuguese Foundation for Science and Technology (FCT) under award numbers CERN/FP/123610/2011 and PTDC/FIS-NUC/1525/2014. University College London and Lawrence Berkeley National Laboratory thank the U.K. Royal Society for travel funds under the International Exchange Scheme (IE141517). 

We acknowledge additional support from the Boulby Underground Laboratory in the U.K., the University of Wisconsin for grant UW PRJ82AJ and the GridPP Collaboration, in particular at Imperial College London. Part of the research described in this Letter was conducted under the Ultra Sensitive Nuclear Measurements Initiative at Pacific Northwest National Laboratory, which is operated by Battelle for the U.S. Department of Energy under Contract No. DE-AC05-76RL01830.

We acknowledge many types of support provided to us by the South Dakota Science and Technology Authority, which developed the Sanford Underground Research Facility (SURF) with an important philanthropic donation from T. Denny Sanford as well as support from the State of South Dakota. SURF is operated by Lawrence Berkeley National Laboratory for the DOE, Office of High Energy Physics. The University of Edinburgh is a charitable body, registered in Scotland, with the registration number SC005336. 
\newpage 

\bibliographystyle{apsrev4-2}
\bibliography{LZNew}

\end{document}